\documentclass[aip,sd,amsmath,amssymb,reprint]{revtex4-1}

\usepackage{graphicx}
\usepackage{dcolumn}
\usepackage{bm}
\usepackage{color}
\begin{document}

\title{Suppression of neuronal phase synchronisation in cat cerebral cortex}

\author{Ewandson L. Lameu}
\author{Fernando S. Borges}
\author{Rafael R. Borges}
\affiliation{P\'os-Gradua\c c\~ao em Ci\^encias, Universidade Estadual 
de Ponta Grossa, Ponta Grossa, PR, Brazil.}
\author{Kelly C. Iarosz}
\author{Iber\^e L. Caldas}
\affiliation{Instituto de F\'isica, Universidade de S\~ao Paulo, S\~ao Paulo, 
SP, Brazil.}
\author{Antonio M. Batista$^a$}
\affiliation{Departamento de Matem\'atica e Estat\'istica, Universidade
Estadual de Ponta Grossa, Ponta Grossa, PR, Brazil.}
\email{antoniomarcosbatista@gmail.com}
\author{Ricardo L. Viana}
\affiliation{Departamento de F\'isica, Universidade Federal do Paran\'a, 
Curitiba, PR, Brazil.}
\author{J\"urgen Kurths}
\affiliation{Department of Physics, Humboldt University, Berlin, Germany; 
Institute for Complex Systems and Mathematical Biology, Aberdeen, Scotland; 
and Potsdam Institute for Climate Impact Research, Potsdam, Germany.}

\date{\today}

\begin{abstract}
We have studied effects of perturbations on the cat cerebral cortex. 
According to the literature, this cortex structure can be described by a 
clustered network. This way, we construct a clustered network with the same 
number of areas as in the cat matrix, where each area is described as a 
sub-network with small-world property. We focus on the suppression of 
neuronal phase synchronisation considering different kinds of perturbations. 
Among the various controlling interventions, we choose three methods: 
delayed feedback control, external time-periodic driving, and activation of 
selected neurons. We simulate these interventions to provide a procedure 
to suppress undesired and pathological abnormal rhythms that can be 
associated with many forms of synchronisation. In our simulations, we have 
verified that the efficiency of synchronisation suppression by delayed feedback 
control is higher than external time-periodic driving and activation of 
selected neurons for the cat cerebral cortex with the same coupling strengths.
\end{abstract}

\maketitle

\begin{quotation}
The cat corticocortical network is organised into visual, auditory, 
somatosensory-motor, and frontolimbic regions, and can be separated into 65 
cortical areas. The areas connected by fibres of different densities that 
can be described through a connection matrix. In this work, we use the cat 
matrix considering that each area corresponds to a small-world sub-network. We 
build a clustered network where the local connections in the small-world 
sub-network are given by electrical synapses, while the shortcuts and the 
connections among the areas are described by chemical synapses. By means of 
this clustered network, we study the suppression of neuronal phase 
synchronisation. Synchronisation might be behind the way we perceive objects, 
but it is also responsible for abnormal behaviour. Clinical evidences pointed 
out that synchronisation of a small group of neurons plays a key role in some 
pathological conditions like Parkinson's disease, tremor, and epilepsy. For 
this reason, the study of control of undesirable neuronal rhythms is relevant 
to restore normal spiking activity in a neuronal network.  We focus on 
three methods of intervention: delayed feedback control, external time-periodic 
driving, and activation of selected neurons. Methods of intervention have 
importance for the treatment of severe neurological and psychiatric diseases.
We have observed that neuronal phase synchronisation can be suppressed by 
means of these methods. Morever, with regard to suppression, we have also 
verified that delayed feedback control has a better efficiency than external 
time-periodic driving and activation of selected neurons.  Feedback control is a
method that has been implemented in clinical applications.
\end{quotation}

\section{Introduction}

The cerebral cortex is an important part of the mammalian brain,  and it is the
outer covering of gray matter over the brain´s hemispheres. It is responsible 
for cognitive tasks like emotion, complex thought, memory, language 
comprehension, and consciousness \cite{buzsaki06}. In the literature, it is 
possible to find information about the structure of the cerebral cortex network 
for the macaque monkey \cite{lund93}, C. elegans \cite{izquierdo13}, the 
cat \cite{scannel95}, etc.  The cat cerebral cortex connectivity data was first 
published by Scannell and Young \cite{scannel93}. The cat corticocortical 
network can be separated into 65 areas, and the areas are connected by fibres 
of different densities that can be described thought a connection matrix. The 
areas are separated into four clusters or cognitive regions named as visual, 
auditory, somatosensory-motor, and frontolimbic. 
 
We have studied a model of neuronal network that contains a connectivity 
configuration in accordance with the cat matrix \cite{gardenes10}. The
clusters formed by cortical areas with common functional roles are responsible 
for the complexity of the cat network \cite{hilgetag00,zhou06,zemanova06}. 
With this in mind, we consider that each element of the matrix has a sub-network 
with small-world property \cite{watts98}. Small-world network is clustered like 
regular networks wich has a path length comparable to random networks. The 
short path length is due to the existence of long-range connections. Stam 
and collaborators \cite{stam07} presented studies about the presence of 
small-world characteristics in functional brain networks, as well as they also 
presented that Alzheimer's disease is characterised by a loss of small-world 
property. Epilepsy in small-world network was investigated by Netoff and 
collaborators \cite{netoff04}. They modelled activity in hippocampal slices 
considering small-world networks of excitatory neurons that reproduce bursts and 
seizures.

In recent years, mathematical models to describe neuronal networks have been 
studied very intensively. Neuronal mathematical models can take many forms, 
e.g. differential equations such as the Hodgkin-Huxley model 
\cite{hodgkin52} and Hindmarsh-Rose model \cite{hindmarsh82}, as well as models
with discrete time as the Rulkov map \cite{rulkov01}. In this work, we consider
a coupled Rulkov map network that presents not only discrete time, but also 
discrete space. The Rulkov map presents two different time scales, where
the variable with slow dynamics is responsable for the modulation of bursts
in the fast variable \cite{dhamala04}. Networks of coupled Rulkov maps have 
been used in studies about neuronal phase synchronisation 
\cite{batista12,lameu16}, suppression of bursting synchronisation 
\cite{lameu12}, and pattern formation \cite{wang07}.
 
Neuronal synchronisation can be found in neuronal activities due to coupling 
among neurons or by means of common inputs.  Studies have demonstrated the 
importance of synchronisation of oscillatory phases between different brain 
regions in memory processes \cite{fell11}.  Roelfsema et al. \cite{roelfsema97} 
realised studies recording local field potentials from electrodes implanted in the 
cortex of cats. They verified large-scale synchronisation when cats were 
submitted to a sudden change of a visual pattern. Synchronisation in 
ensembles was also studied by Ivanchenko et al. \cite{ivanchenko04}, who 
observed a second-order phase transition to synchronisation. Axmacher et al. 
\cite{axmacher06} verified that specific forms of cellular plasticity during 
subsequent stages of memory formation are induced by synchronisation. 
Nevertheless, there are evidences that certain brain disorders are related to 
neuronal synchronisation, e.g. epilepsy that results from high and 
extended synchronisation \cite{uhlhaas06}. Moreover, synchronisation of 
neuronal activity is a common finding in patients with Parkinson's disease 
\cite{chen07}. Levy et al. \cite{levy00} demonstrated  that Parkinsonian 
patients present synchronised high-frequency activity in the subthalamic 
nucleus.

In face of neuronal synchrony in brain disorders, we study here three methods
to suppress undesired synchronisation: delayed feedback control, external 
time-periodic driving, and activation of selected neurons. The delayed 
feedback control was proposed by Rosenblum and Pikowsky \cite{rosenblum2004} to
suppress synchronised pathological brain rhythms through delayed feedback signal
\cite{batista10}. Batista et al. \cite{batista13} observed that 
feedback control in networks of Hodgkin-Huxley-type neurons with chemical 
synapses can present more energy saving when compared to other suppression 
methods. Feedback control is a method that has been implemented in clinical 
applications by means of functional magnetic resonance imaging based 
neurofeedback \cite{linden12}. With regard to external time-periodic driving, 
it was verified that electrical  stimulation of deep brain structures can reduce or 
completely suppresses seizures \cite{lesser03}. In addition, the method about 
activation of selected neurons was used by Han and Boyden \cite{han07}. They 
used light pulses on genetically targeted neurons not only for activation, but also 
for inhibition of neuronal activity.

All in all, we consider a clustered network that is composed by the
matrix of corticocortical connections in the cat with small-world sub-networks
using as local dynamics a two-dimensional map to describe the neuronal
activity. We build small-world sub-networks according to the procedure 
proposed by Newman and Watts \cite{newman99}, who inserted randomly chosen 
shortcuts in a regular network. With the objective of finding an effective way 
of suppression of neuronal synchronisation, we study 3 methods:
external time-periodic signal, neuron control with light, and time-delayed 
feedback signal. One of our main numerical results is to show that the 
time-delayed feedback signal is more effective than the control with light and 
feedback signal in the cat cerebral cortex. We also verify that a perturbation 
in the frontolimbic region affects the other regions.

The paper is organised as follows: In Section 2, we present the mathematical
model of the neuronal network. Section 3 shows the burst phase
synchronisation. Section 4 exhibits our numerical results with the different 
methods of suppression of neuronal phase synchronisation. In the last
Section, we draw our conclusions.

%%%%%%%%%%%%%%%%%%%%%%%%%%%%%%%%%%%%%%%
%%%%%%%%%%%%%%%%%%%%%%%%%%%%%%%%%%%%%%%

\section{Clustered network of Rulkov neurons}

\begin{figure}[hbt]
\centering
\includegraphics[height=5cm,width=8cm]{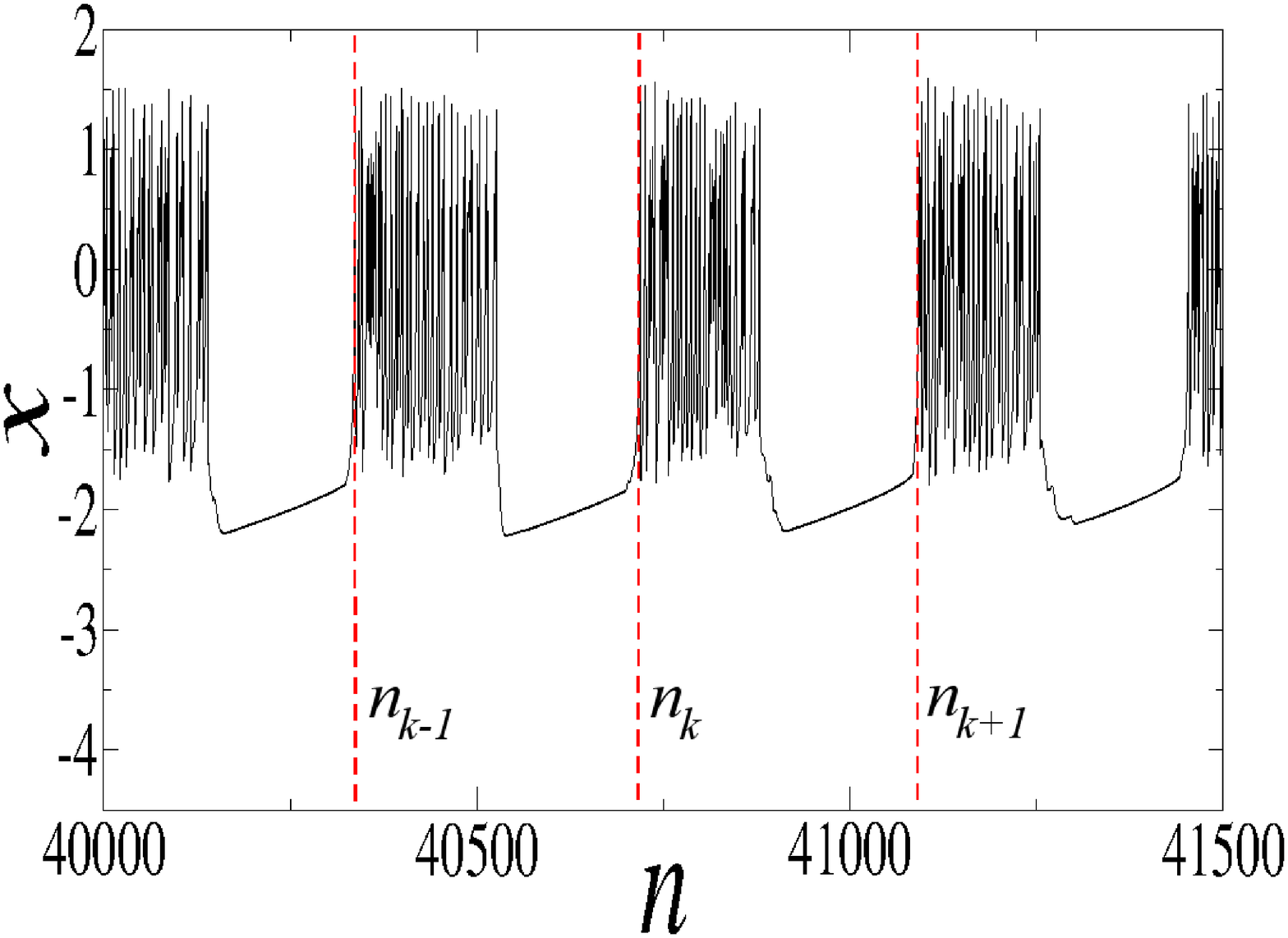}
\caption{Time evolution of the fast variable in the Rulkov map, where $n_k$ 
denotes when a new neuronal bursting starts, and $k$ is an integer}
\label{fig1}
\end{figure}

We consider, as neuronal model, the Rulkov map \cite{rulkov01} that reproduces
neuronal bursting by means of two variables, and is given by
\begin{eqnarray}
x_{n+1} &=& \frac{\alpha}{1 + x_n^2} + y_n, \\
y_ {n+1} &=& y_n - \sigma(x_n - \rho),
\end{eqnarray}
where $x_n$ is the fast dynamical variable, $y_n$ is the slow dynamical 
variable, $\alpha$ controls the duration of bursts, $\sigma$ and $\rho$ 
describe the slow time-scale. Figure \ref{fig1} exhibits irregular sequence
of bursts of the fast variable, where $n_k$ denotes when the neuronal bursting
starts, and $k$ is an integer.

\begin{figure}[hbt]
\centering
\includegraphics[height=8cm,width=8cm]{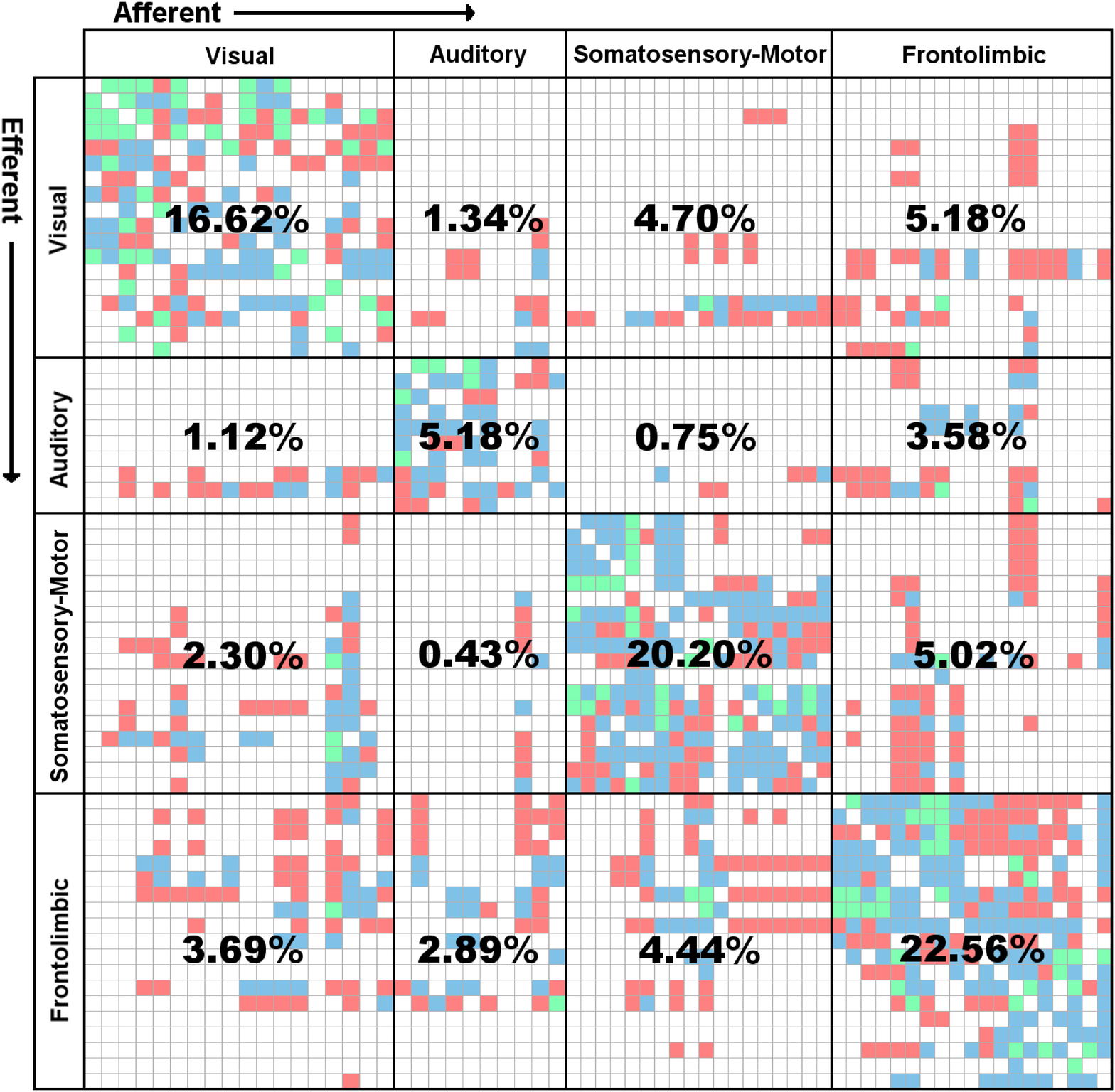}
\caption{(Colour online) Density of connections between cortical areas 
classified as absent of connection (white), sparse (red), intermediate (blue), 
and dense (green). The percentages correspond to the amount of intra and 
inter connections in relation to the total connectivity matrix.}
\label{fig2}
\end{figure}

In accordance with the matrix, that describes the corticocortical connectivity
of the cat brain, obtained by Scannel et al. \cite{scannel95}, we build a 
clustered neuronal network. Figure \ref{fig2} shows the densities of 
connections by means of colours, where we can see white for no connections, 
sparse connections in red, intermediate connections in blue, and dense 
connections in green. Each one of the $65$ areas is modelled by a 
small-world network with $100$ neurons and $5\%$ of shortcuts. The cortical 
areas classified as sparse (red) have $50$ randomly directed connections. The 
intermediate connectivities (blue) and  the dense connectivities (green) have
$100$ and $150$ randomly directed connections, respectively.

The areas are separated into 4 cognitive regions: visual, auditory, 
somatosensory-motor, and frontolimbic. The visual region is composed of 18 
cortical areas and the auditory of 10 areas, while the somatosensory-motor and 
the frontolimbic present 18 and 19 cortical areas, respectively. The values of 
percentages, in Figure \ref{fig2}, describe the amount of connections 
between regions in relation to the total connectivity in the matrix. For 
instance, inside the visual region there is $16.62\%$ of the total 
connectivity, and the percentage of connections to the visual region from the
auditory region is equal to $1.34\%$. The frontolimbic region has the largest 
amount of connections in the cat matrix.

In this work, the dynamics of the clustered network is based on the cat matrix 
and given by \cite{lameu16}
\begin{eqnarray}\label{eqcoupling}
x_{n+1}^{(i,p)} & = & \frac{\alpha^{(i,p)}}{1+(x_n^{(i,p)})^2}+y_n^{(i,p)} 
+\frac{g_e}{2}(x_n^{(i-1,p)}+x_n^{(i+1,p)} \nonumber \\  
& & -2x_n^{(i,p)})-g_c\sum_{d=1}^S\sum_{f=1}^P\left[ A_{(d,f),(i,p)}H(x_n^{(d,f)}
\right.  \nonumber \\  
& &  \left. -\theta)(x_n^{(i,p)}-V_s)\right ] +\Lambda_n, \\
y_{n+1}^{(i,p)} & = & y_n^{(i,p)}-\sigma(x_n^{(i,p)}-\rho),
\end{eqnarray}
where $(i,p)$ denotes the neuron $i$ ($i=1,2,\dots,S$) in the cortical 
area $p$ ($p=1,2,\dots,P$), $S=100$ and $P=65$ are the quantity of neurons
in each small-world sub network and the number of cortical areas, respectively.
The third term of the first equation corresponds to the electrical coupling 
with strength $g_e$ and the fourth term is the chemical coupling with strength
$g_c$. In the chemical coupling term, the chemical connection between one
neuron ($i,p$) and other neuron ($d,f$) is given by the adjacency matrix 
$A_{(d,f),(i,p)}$. In addition, $H(x)$ is the Heaviside step function, where 
$\theta=-1.0$ is the presynaptic threshold for the chemical synapse, and
$V_s$ is the reversal potential. In our simulations, we have considered that
$\alpha^{(i,p)}$ is randomly distributed in the interval $[4.1, 4.4]$, 
$\sigma=0.001$, $\rho=-1.25$, $\theta=-1.0$, $V_s=1.0$ for excitatory 
synapses, $V_s=1.0$, $V_s=-2.0$ for inhibitory, and 3 different methods for 
the perturbation $\Lambda_n$. With regard to the synapses, we consider that
the electrical synapses are the connections between the nearest neighbours 
into the small-world networks, while the chemical synapses are  the shortcut
connections into the small-world and connections between areas. We have
also consider $75\%$ of excitatory and $25\%$ of inhibitory chemical
synapses in all the network.

\section{Neuronal phase synchronisation}

In this work, we calculate the neuronal phase by means of the time evolution 
within each burst, varying from 0 to $2\pi$ as $n$ evolves from $n_k$ to 
$n_{k+1}$ (Fig. \ref{fig1}),
\begin{equation}
\phi_n=2\pi k+2\pi \frac{n-n_k}{n_{k+1}-n_k}.
\end{equation}
Through the phase we compute the Kuramoto's order parameter $R_n$ to 
check the synchronous behaviour, that is given by
\begin{equation}\label{porder}
z_n^{(l)}=R_n^{(l)}\exp ({\rm i}\Phi_n^{(l)})\equiv 
           \frac{1}{N_{l}}\sum_{j\in I_{l}}\exp ({\rm i}\phi_n^{(j,I_l)}),
\end{equation}
where $R_n$ and $\Phi_n$ are the amplitude and  the angle of a centroid phase 
vector, respectively. $I_{l}$ denotes the cognitive areas, with $l=1$ for 
visual, $l=2$ for auditory, $l=3$ for somatosensory-motor, and $l=4$ for
frontolimbic. $N_{l}$ corresponds to the number of neurons of each area. 
The phase of the neurons $j$, in the cortical area $I_l$, is denoted by
$\phi_n^{j,I_l}$. The order parameter is equal to 1 for a completely synchronous
behaviour and much less than 1 for uncorrelated phases.

The phase synchronisation as a function of the coupling strength can be 
analysed by means of the time average order parameter, given by 
\begin{equation}
{\bar R}=\frac{1}{n_{\rm final}-{n_{\rm initial}}}
\sum_{n_{\rm initial}}^{n_{\rm final}} R_n ,
\end{equation} 
where $n_{\rm final}-n_{\rm initial}$ is the time window for measurements.

\begin{figure}[hbt]
\centering
\includegraphics[height=5cm,width=8cm]{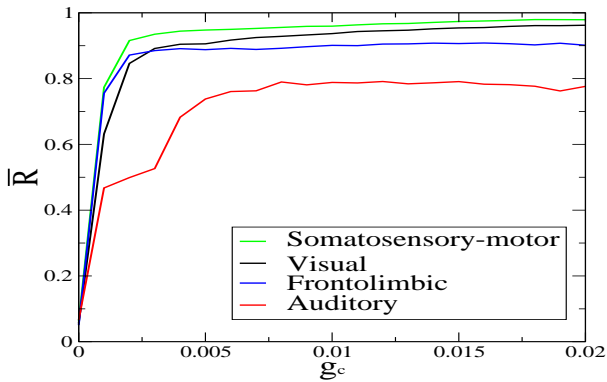}
\caption{(Colour online) Average order parameter of the cat's cognitive brain 
regions as a function of the chemical coupling strength $g_c$ for a fixed 
electrical coupling $g_e=0.05$. We consider $50,000$ iterations where the
first $20,000$ were excluded as transient.}
\label{fig3}
\end{figure}

Figure \ref{fig3} shows the time average order parameter as a function of the
chemical coupling strength for each cortical region, where the value of the 
electrical coupling $g_e=0.05$. We can see that the values of saturation of 
${\bar R}$ for the visual and the somatosensory-motor regions are larger 
than those of the auditory and frontolimbic regions. The auditory region
presents a small percentage of intra connections, and due to this fact it is
not possible to observe a strong synchronised state. Consequently, the time 
average order parameter saturates in a value smaller than $0.8$. In addition, 
we also observe that all transitions are of second-order.
 
\section{Suppression of neuronal phase synchronisation}

The neuronal synchronisation can be associated with brain disorders, such as 
epilepsy and Parkinson's disease. Due to this fact, we have studied methods of
suppression of neuronal phase synchronisation. As diagnostic tool of suppression
we use the suppression factor \cite{rosenblum04}, given by  
\begin{equation}
S=\sqrt{\frac{{\rm Var}(M_n)}{{\rm Var}(M_n^p)}},
\end{equation}
where ${\rm Var}()$ is the variance, $M_n$ and $M_n^p$ are the mean fields 
of the fast dynamical variable $x_n$ in the absence and presence of the control, 
respectively. The variance of $M_n^p$ is small when the synchronisation is 
suppressed, consequently the suppression factor $S$ is strongly increasing. 
Nevertheless, $S$ has a value approximately equal to one when the control 
is not efficient to suppress the synchronisation.

We have considered three methods of suppression that are found in literature:
delayed feedback control, external time-periodic driving, and activation of
selected neurons. 
\begin{itemize}
\item[(i)] In the delayed feedback control, the last term in eq. 
(\ref{eqcoupling}) has the following form
\begin{equation}
\Lambda_n=\frac{1}{N_l}\sum_{(i,p)\in I_l}x_{n-\tau}^{(i,p)},
\end{equation}
where $\tau$ represents the number of interations before, and the control is 
only applied in one of the four cognitive areas in $N_l=100$ neurons randomly 
chosen at each interation. 
\item[(ii)] With regard to external time-periodic driving, we consider
\begin{equation}
\Lambda_n=I \sin(\omega n),
\end{equation}
where $I=1$ and $\omega=1$ are the amplitude and frequency of the perturbation,
respectively. The time-periodic driving is also applied on the 100 neurons 
randomly chosen in the cognitive area. 
\item[(iii)] The third method is similar to light stimulation, in other words, the 
neuron receives a light pulse and goes to a state where it is forced to spike. In 
our simulations, $100$ randomly chosen neurons are activated ($x_n=1.2$) in a 
specific cognitive area when the perturbation is applied.
\end{itemize}

\begin{figure}[hbt]
\centering
\includegraphics[height=9cm,width=7cm]{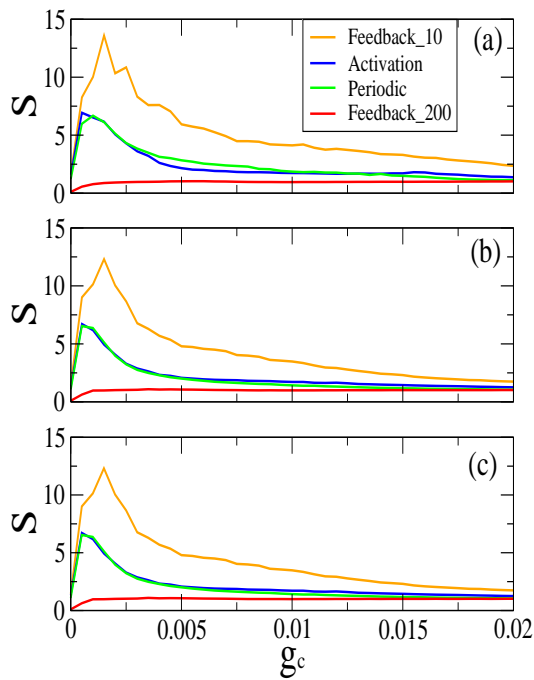}
\caption{(Colour online) Suppression factor $S$ as a function of the chemical
coupling strength $g_c$ for $g_e=0.05$ and the different controls applied on 
(a) the visual, (b) the auditory, and (c) the somatosensory-motor areas.}
\label{fig4}
\end{figure}

In Figure \ref{fig4}, the suppression factor is calculated for the controls
applied on (a) the visual, (b) the auditory, and (c) the somatosensory-motor 
regions varying the chemical coupling strength. The feedback is represented by 
orange and red lines for $\tau$ equal to $10$ and $200$ iterations, 
respectively. The external time-periodic driving is denoted by green line, 
whereas the activation of selected neurons is represented by blue line. Our 
results show that the feedback control with $\tau$ equal to $200$ does not 
produce a suppression of the synchronisation. Conversely, for $\tau=10$, for 
activation, and for time-periodic driving it is possible to observe a suppression 
with $S>2$ for a chemical coupling strength smaller than $0.01$. The 
activation and the time-periodic driving have approximately the same behaviour 
of the suppression factor. We also verify that the feedback with $\tau=10$ 
presents the largest value of suppression factor. In addition, the other cognitive 
areas do not suffer a significant effect from these controlled areas (visual, 
auditory, and somatosensory-motor), in other words, the unperturbed areas 
remain in a synchronised state.

\begin{figure}[hbt]
\centering
\includegraphics[height=8.5cm,width=8.5cm]{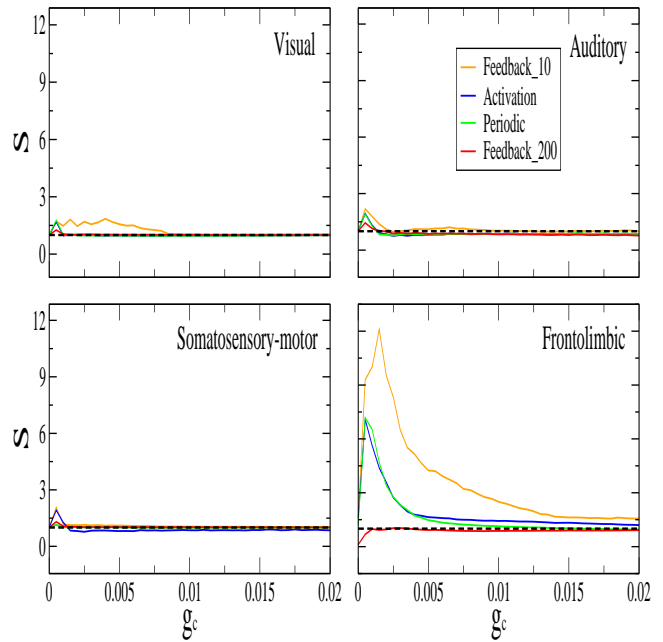}
\caption{(Colour online) Suppression factor as a function of the chemical
coupling strength for (a) the visual, (b) the auditory, (c) the 
somatosensory-motor, and (d) frontolimbic areas, where the controls are 
applied on the frotolimbic area and $g_e=0.05$.}
\label{fig5}
\end{figure}

In the same way that the controls are applied on the visual, the auditory, and
the somatosensory-motor, we apply on the frontolimbic area. Unlike the before
cases, the uncontrolled cognitive areas presents a small influence from the
suppression in the frontolimbic area. The suppression factor has its maximum value
approximately equal to $2$ in a small range of the chemical coupling strength,
as shown in Figure \ref{fig5} for the visual, auditory, and somatosensory-motor
areas. In the frontolimbic area, the values of the suppression factor have a
behaviour similar to that obtained in Figure \ref{fig4}. In fact, the feedback
with $\tau$ equal to $10$ presents the best efficiency compared with the
activation, the time-periodic driving, and with $\tau$ equal to $200$.

\begin{figure}[hbt]
\centering
\includegraphics[height=8.5cm,width=8.5cm]{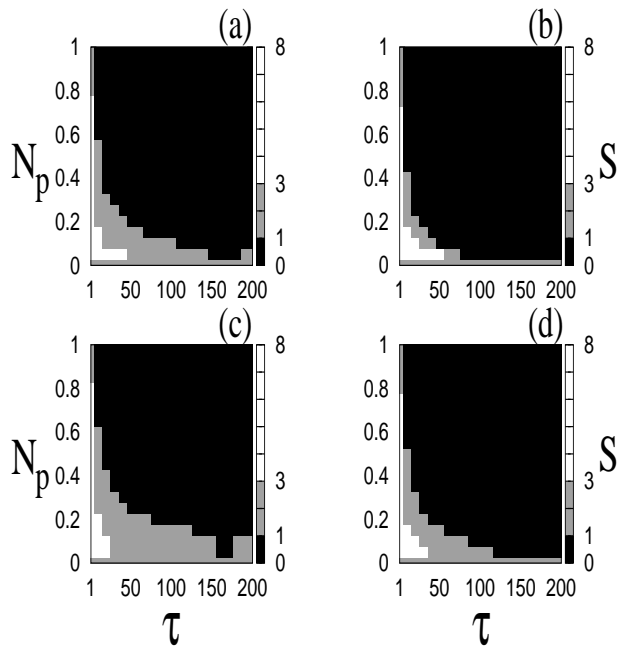}
\caption{Suppression factor in the bar for the percentage of perturbed 
neurons $N_p$ versus the time delay $\tau$, where we consider $g_e=0.05$ 
and $g_c=0.005$. Delayed feedback control applied on (a) the visual, (b) the 
auditory, (c) the somatosensory-motor, and (d) the frontolimbic areas.}
\label{fig6}
\end{figure}

In view of our results from the delayed feedback control better than the
activation and time-periodic driving, we analyse how the suppression factor is
affected by the time delay $\tau$ and the percentage of perturbed neurons 
$N_p$. In Figure \ref{fig6} we can see the suppression factor in the bar for 
the number of perturbed neurons versus the time delay, where the delayed 
feedback is applied on (a) the visual, (b) the auditory, (c) the 
somatosensory-motor, and (d) the frontolimbic areas. The black regions 
correspond to the case in which the cognitive area does not present a 
suppression of synchronisation ($S\leq 1$), the grey regions exhibit a small 
suppression of synchronisation with the suppression factor in the interval 
$1<S\leq 3$, and the white regions show the values of $N_p$ and $\tau$ 
where the feedback delayed control is more efficient ($S>3$). As a result, 
we verify that the method by means of feedback delayed control is efficient 
not only for different $N_p$ values, but also for small time delay.

The time evolution of the mean field for the black region in Figure
\ref{fig6} presents a large amplitude oscillations, as a result of the
synchronised behaviour. Whereas, in the white region, where there is
no evidence of synchronisation, the mean field has small amplitude 
oscillations. Oscillation quenching has been investigated in systems
of coupled nonlinear oscillators \cite{koseska13}, that are classified
in: oscillation (OD) and amplitude death (AD). In our simulations,
we have verified that the 3 methods of suppression can induce to a
mean field amplitude death (MFAD). This way, $S$ can be used
as a disgnostic tool of MFAD.

\section{Conclusions}

In this work, we studied suppression of burst phase synchronisation in a 
neuronal network with a structure according to the corticocortical connections 
of the cat brain. We have considered the cat matrix, where each cortical area
has a sub-network with small-world properties.

Considering an initial configuration in that the neuronal network presents a 
synchronous behaviour, we have applied and compared three different 
suppression methods: delayed feedback control, external time-periodic 
driving, and activation of neurons. As a result, we verify that it is possible to 
obtain suppression by means of the three methods. The methods produce 
suppression only in the cortical areas where are applied, except when they 
are applied on the frontolimbic area. We observe a small suppression in the 
other areas when the perturbations are applied on the frontolimbic area. 
This occurs due to the fact that the frontolimbic area presents a larger 
external connectivity than the other areas.

In our simulations, using the suppression factor as diagnostic tool, the 
delayed feedback control have shown the best efficiency compared with the
external time-periodic driving and the activation of neurons. In addition,
we have verified that the delayed feedback is better for small values of the
time-delayed in a large range of the number of controlled neurons. The delayed
feedback control does not damage the neurons due to the fact that it uses a
signal amplitude presented by neuronal activity. 

%%%%%%%%%%%%%%%%%%%%%%%%%%%%%%%%%%%%%
%%%%%%%%%%%%%%%%%%%%%%%%%%%%%%%%%%%%%

\begin{acknowledgments}
We wish to acknowledge the support of the Brazilian agencies: CNPq, CAPES,
and FAPESP (2015/07311-7 and 2011/19296-1).
\end{acknowledgments}

\end{document}